\documentclass[12pt,preprint,twocolumn]{aastex631}
\usepackage{amssymb,amsmath,graphicx,calc}
\setcitestyle{notesep={ }}

\begin{document}

\title{Disk Wind Feedback from High-mass Protostars. II. The Evolutionary Sequence}
\author[0000-0001-9040-8525]{Jan E. Staff}
\affil{Department of Space, Earth \& Environment, Chalmers University of Technology, Gothenburg, Sweden\\ and\\ College of Science and Math, University of the Virgin Islands, St
Thomas, 00802, United States Virgin Islands}
\author[0000-0002-6907-0926]{Kei E. I. Tanaka}
\affil{Department of Earth and Planetary Sciences, Tokyo Institute of Technology, Meguro, Tokyo, 152-8551, Japan,\\and\\
Center for Astrophysics and Space Astronomy, Department of Astrophysical and Planetary Sciences, University of Colorado Boulder, Boulder, CO 80309, USA,
ALMA Project, National Astronomical Observatory of Japan, Mitaka, Tokyo 181-8588, Japan}
\author[0000-0002-3835-3990]{Jon P. Ramsey}
\affil{Department of Astronomy, University of Virginia, Charlottesville, VA 22904, USA}
\author[0000-0001-7511-0034]{Yichen Zhang}
\affil{Department of Astronomy, University of Virginia, Charlottesville, VA 22904, USA}
\author[0000-0002-3389-9142]{Jonathan C. Tan}
\affil{Department of Space, Earth \& Environment, Chalmers University of Technology, Gothenburg, Sweden\\ and\\ Department of Astronomy, University of Virginia, Charlottesville, VA 22904, USA}

\date{}

\begin{abstract}
Star formation is ubiquitously associated with the ejection of accretion-powered outflows that carve bipolar cavities through the infalling envelope. This feedback is expected to be important for regulating the efficiency of star formation from a natal pre-stellar core. These low-extinction outflow cavities greatly affect the appearance of a protostar by allowing the escape of shorter wavelength photons. Doppler-shifted CO line emission from outflows is also often the most prominent manifestation of deeply embedded early-stage star formation. Here, we present 3D magneto-hydrodynamic simulations of a disk wind outflow from a protostar forming from an initially $60\:M_\odot$ core embedded in a high pressure environment typical of massive star-forming regions. We simulate the growth of the protostar from $m_*=1\:M_\odot$ to $26\:M_\odot$ over a period of $\sim$100,000 years. The outflow quickly excavates a cavity with half opening angle of $\sim10^\circ$ through the core. This angle remains relatively constant until the star reaches $4\:M_\odot$. It then grows steadily in time, reaching a value of $\sim 50^\circ$ by the end of the simulation. We estimate a lower limit to the star formation efficiency (SFE) of 0.43. However, accounting for continued accretion from a massive disk and residual infall envelope, we estimate that the final SFE may be as high as $\sim0.7$. We examine observable properties of the outflow, especially the evolution of the cavity opening angle, total mass and momentum flux, and velocity distributions of the outflowing gas, and compare with the massive protostars G35.20-0.74N and G339.88-1.26 observed by ALMA, yielding constraints on their intrinsic properties.
\end{abstract}

\section{Introduction}
\label{introsection}

Low-mass stars and their associated accretion disks form from gravitationally bound cores \citep{1987ARA&A..25...23S} and are frequently associated with the launching of bipolar jets and outflows \citep[for reviews, see, e.g.,][]{2014prpl.conf..451F,2016ARA&A..54..491B}.
The magnetocentrifugal mechanism \citep{1982MNRAS.199..883B,1983ApJ...274..677P,2000prpl.conf..759K} is widely thought to be responsible for launching, accelerating and collimating these protostellar outflows.
In this scenario, the combination of large-scale magnetic fields with gravity and rotation results in the ejection, acceleration and then collimation of gas originating from the surface of the accretion disk. 
A number of numerical simulation studies have been performed to investigate this process across a variety of different conditions and assumptions \citep[e.g.,][]{1985PASJ...37...31S,1985PASJ...37..515U,2003ApJ...582..292O,1997Natur.385..409O,1997ApJ...482..712O,1997ApJ...482..708R,1999ApJ...526..631K,2011ApJ...728L..11R,2010ApJ...722.1325S,2015MNRAS.446.3975S,2019ApJ...882..123S,2006ApJ...653L..33A,2007A&A...469..811Z,2012ApJ...746...96T,2012ApJ...757...65S,2014MNRAS.439.3641S,2014ApJ...793...31S,2016ApJ...825...14S,2019MNRAS.484.2364R,2020ApJ...896..126G,2020ApJ...900...59M,2020ApJ...900...60M}.
However, alternative theoretical scenarios have also been proposed as being relevant for outflow launching, including: the X-wind model involving the interaction of the protostellar magnetic field with the inner disk \citep[e.g.,][]{1991ApJ...379..696L,2000prpl.conf..789S}; stellar wind driven outflows \citep[e.g.,][]{2005ApJ...632L.135M}; and magnetic pressure driven outflows \citep{1996MNRAS.279..389L}. 

Observationally, support for the disk wind model in low- and intermediate-mass systems has been provided by high angular resolution observations of a handful of systems, e.g., TMC1A \citep{2016Natur.540..406B}, HH212 \citep{2017NatAs...1E.152L}, DG Tau B \citep{2020A&A...634L..12D}, and IRAS 21078+5211 \citep{2022NatAs...6.1068M}. In each case, the launching of the outflow can be traced to the accretion disk, demonstrating a launching radius that extends out to scales of up to $\sim$ 20 au from the central protostar.

The formation of high-mass stars is more difficult to characterize observationally as there are fewer sources, they are farther away and they are more obscured by surrounding gas and dust.
Nevertheless, massive star formation is also typically observed to be associated with the launching of bipolar jets and outflows \cite[see, e.g.,][]{2007prpl.conf..245A,2014prpl.conf..149T,2016A&ARv..24....6B, 2017NatAs...1E.146H}.
For example, the central source powering HH 80 and HH 81 (IRAS 18162-2048) \citep{1993ApJ...416..208M}, and G339.88-1.26 \citep{2019ApJ...873...73Z} are both associated with highly collimated outflows. Another massive protostar, G35.20-0.74N, has also been found to launch a highly collimated jet \citep[e.g.,][]{2019NatCo..10.3630F}.
Indeed, \citet{2015A&A...573A..82C} found that outflows from a number of intermediate and high-mass protostars appear as scaled-up versions of those from low-mass protostars, while \citet{2020ApJ...904..139S} also found this to be the case for the outflow from the massive protostar NGC 7538 IRS1.
Wider angle molecular outflows have also been observed from massive protostars \citep[e.g.][]{2002A&A...383..892B,2004A&A...426..503W,2013ApJ...767...58Z,2014ApJ...792..116Z,2015MNRAS.453..645M}.
In general, the trend is that higher luminosity, i.e., more massive, protostars tend to have more powerful and more massive outflows with wider opening angles than their low-mass counterparts.

\citet{2002Natur.416...59M} suggested that a combination of turbulence and magnetic pressure provides most of the support in a massive pre-stellar core against gravity.
In this ``Turbulent Core Accretion'' (TCA) model, high-mass star formation is a scaled-up version of low-mass star formation, with accretion rates expected to be $\sim10^{-4}$ to $\sim 10^{-3}\:M_\odot\:{\rm yr^{-1}}$, compared to $\sim10^{-6}$ to $\sim 10^{-5}\:M_\odot\:{\rm yr^{-1}}$ in lower-mass cores. If that is the case, then outflows from forming massive stars can therefore also be a scaled-up version of the outflows from lower-mass forming stars.

Other formation scenarios for high-mass stars have also been proposed. \citet{1998MNRAS.298...93B} suggested that high-mass stars form by the collision of multiple smaller objects that formed close together.
Another possibility suggested by \citet{2001MNRAS.323..785B} is that massive stars form together in the central region of dense proto-clusters, where most of the mass is accreted from a globally collapsing clump \citep[see][for a review of these scenarios]{2014prpl.conf..149T}.
This could lead high-mass stars to accrete from smaller disks that change orientation over time, leading to outflows that also keep changing directions \citep{2020ApJ...905...25G}.

In contrast to outflows from low-mass protostars, it is still debated whether or not strong magnetization is required to drive an outflow from high-mass protostars.
For example, \citet{2020MNRAS.499.4490M} found that, in their simulations, the outflow launching failed or was much delayed unless the initial cloud was strongly magnetized. 
In contrast, \citet{2020ApJ...904..168B}, based on observations, argued for a weak magnetization in the case of G327.3, despite it also having an outflow. 
The direction of the magnetic field in the core is also debated; in some cases, it has been found to be parallel to the outflow and perpendicular to the disk \citep{2010Sci...330.1209C, 2015A&A...583L...3S}, while other studies have found that the outflow axis is randomly oriented with respect to the core-field \citep{2014ApJ...792..116Z}. From an analysis of about 200 outflows, \citet{2022ApJ...941...81X} find evidence for preferential alignment of outflow directions with large-scale $B-$fields, but with significant scatter for any given outflow to $B-$field to orientation.

\citet{2019ApJ...882..123S} (hereafter Paper I) presented 3D magneto-hydrodynamic (MHD) simulations of disk wind outflows from a $60\:{M_\odot}$ core, but with the protostellar mass, accretion rate and mass outflow rate held at fixed values representing various stages of the protostellar evolution. The method was to run each simulation for a roughly a local accretion time to infer the properties of the outflow cavity - envelope system. However, because of this approximation this method involved significant uncertainties.

In this paper, i.e., Paper II, we present MHD simulations similar to Paper I, but importantly we now evolve the simulation continuously, following the protostellar evolutionary sequence consistently as the mass of the protostar grows from $m_*=1\:M_{\odot}$ to more than $24~M_{\odot}$. As in Paper I, we start from a $60\:{M_\odot}$ core embedded in a clump with mass surface density of $\Sigma_{\rm cl}=1\:{\rm g\:cm}^{-2}$ within the framework of the Turbulent Core Accretion model of \citet{2002Natur.416...59M,2003ApJ...585..850M}. The protostar increases its mass by accreting material from the envelope through a disk-like boundary condition.
A disk-wind (launched from the accretion disk) is injected into the simulation box, where some envelope material becomes entrained by the outflow.
We simulate the outflow as it propagates through the envelope to investigate the interaction between the wind and the envelope material, and to investigate how much envelope material is displaced, providing us with an estimate of the star formation efficiency. We also compare our simulation results with observations of outflows from massive protostars.

In \S\ref{methodssection} we describe our numerical methods. We present our results in \S\ref{resultssection} and discuss their implications in \S\ref{discussionsection}. Finally, we summarize our findings in \S\ref{summarysection}.

\section{Methods}
\label{methodssection}

The goal of this work is to simulate a magnetically-powered outflow from a massive, growing protostar. 
Using the ZEUS-MP code \citep{2000RMxAC...9...66N}, we conduct a 3D, ideal MHD simulation of an outflow from a massive protostar in the framework of the turbulent core accretion model \citep{2003ApJ...585..850M,2014ApJ...788..166Z,2018ApJ...853...18Z}. In brief, ZEUS-MP solves the equations of magnetohydrodynamics using operator splitting and a staggered mesh, wherein vector quantities (momentum and magnetic field) are offset by half a cell relative to scalar quantities (such as density). Our version of ZEUS-MP uses the Method of Characteristics-Constrained Transport (MOCCT) scheme \citep{1995CoPhC..89..127H}. ZEUS-MP also uses an \citet{1950JAP....21..232V} type artificial viscosity to stabilize the code. In this study, we employed a compressive artificial viscosity coefficient of 2.0 (\texttt{qcon} in the code).
As in Paper I, we consider an initial core of mass of $60\:{M_\odot}$. 
However, in contrast to Paper I, rather than simulating a sequence of separate models for different fixed values of protostellar mass, $m_*$, here we follow the evolution of a single simulation and the resulting outflow for more than 100,000 years as the central star grows from an small initial mass of $m_*=1\:M_\odot$. The setup of the simulation is described below.

\subsection{Simulation domain and boundary conditions}
\label{sub:grid+bcs}

We use a Cartesian grid with $168\times280\times280$ cells in the $x_1$, $x_2$, and $x_3$ directions, respectively, for our ``medium'' resolution simulation. A ``high'' resolution simulation is also run for the earlier phases of the evolution with $336\times560\times560$ cells (see \S\ref{sub:resolution}).
A logarithmic grid (``ratioed'' in ZEUS terminology) is employed, where cells become larger in each direction in a regular fashion as the distance from the origin increases.
This allows us to cover a fairly large spatial region, while maintaining a reasonably high resolution in the central region. The $x_1$ direction (perpendicular to the disk and parallel to the outflow) extends from $100~{\rm au}$ above the disk midplane to $26,500~{\rm au}$, while the $x_2$ and $x_3$ directions (parallel to the disk plane) extend out to $\pm 16,000~{\rm au}$. 
Compared to the Paper I simulations, this domain is about twice as long in the $x_1$ direction, and slightly larger in the $x_2$ and $x_3$ directions.

All boundaries, except for the inner $x_1$ boundary, are outflow boundaries.
The inner $x_1$ boundary is more complicated, as the outflow is injected through it, and mass can ``accrete'' onto the disk through it.
The fastest part of the disk wind is injected in a circular region with radius $r_{\rm inj}$ centered on the origin. As in Paper I, $r_{\rm inj}$ is related to the size of the disk around the protostar, $r_d$ (see eq. 2 of Paper I).
Just outside of the injection region is a smoothing region, through which material is also injected.
The role of this smoothing region is to gradually transition from the density and velocity of the injected disk wind profile to that of the surrounding environment.
The smoothing region has a radius of $r_o=1.8 r_{\rm inj}$, somewhat larger than the value of $r_o=1.3 r_{\rm inj}$ used in Paper I to ensure that it contains several cells (up to 7 cells by the end of simulation) in the $x_2$ and $x_3$ directions at all times.
Going further out is the accretion region, extending from $r_o$ to $r_{\rm acc}$, through which material is removed to join the accretion disk at a controlled rate. Beyond this, we use reflecting boundaries, to prevent any additional mass from flowing off the grid.

The value of $r_{\rm inj}$ (and thus also $r_o$) increases during the evolution as the star grows in mass, since $r_d \propto m_*^{2/3}$ in the fiducial model of \cite{2014ApJ...788..166Z} in the limit of constant star formation efficiency, a fixed disk to star mass ratio, and a constant profile of rotational energy to gravitational energy ratio of material in the initial core.
The radius of the accretion region, $r_{\rm {acc}}$, adjusts over time so that the integrated mass flow rate through the annulus given by $r_{\rm {acc}} - r_o$ that has outflow boundary conditions is $\dot{m}_{\rm sim}=\frac{1}{2}\dot{m}_{*}(1+1/3+1/10)\simeq 0.72\, \dot{m}_{*}$.
Note, the term $1/3$ accounts for the growth of the accretion disk, which is assumed to have a mass $m_{d}= m_*/3$\sout{)}. The term $1/10$ is present to account for the injected mass flux of the disk wind that is immediately returned to the simulation grid through the injection region. The factor $1/2$ is present since we simulate only one hemisphere.
The outer radius of the accretion region, $r_{\rm {acc}}$, is adjusted so that the desired accretion rate is achieved via this region of outflow boundary condition.

\subsection{Initial core}
\label{sub:initialsetup}

We initialize the simulation with a $1\:M_\odot$ protostar located at the origin of our coordinate system, which is 100 au below the inner $x_1$ boundary.
On the grid, we include one hemisphere of a $60\:M_\odot$ core, with a radius of $R_c$ =
12,000 au, which is the size expected for such a core embedded in a clump with mass surface density of $\Sigma_{\rm cl}=1\:{\rm g\:cm}^{-2}$.

In the TCA model, the fiducial initial density structure of the prestellar core is assumed to be spherical, with a power-law of the form $\rho \propto r^{-k_\rho}$ with $k_\rho=3/2$. Thus our density structure is given by
\begin{equation}
\rho(t=0) = \rho_{s} \left( r/R_c \right)^{-3/2},
\end{equation}
where $\rho_s$ is the density at the surface of the core.
Note, in Paper I, which was mainly considering snapshots of later phases of the evolution, we adopted $k_\rho=1$ as an approximation of the expected structure that develops in the expansion wave of the collapse solution.
For our core with $k_\rho=3/2$, we have
$\rho_s = 2.5\times10^{-18}~{\rm g~cm^{-3}}$, i.e., $n_{\rm H}=1.1\times10^{6}\:{\rm cm}^{-3}$ assuming a mass per H of $2.34\times10^{-24}\:{\rm g\:cm}^{-3}$.
Beyond $R_c$ we adopt a constant ambient density of $0.1\rho_s$.
The material in the core and its surroundings is initialized to be at rest.

Following Paper I, the initial magnetic field configuration is the canonical Blandford \& Payne (``BP'') configuration \citep{1982MNRAS.199..883B}, with a constant field added to it to ensure that the core flux is $\sim 1~{\rm mG}\times R_{c}^2$. The BP configuration is a force-free, hour-glass shaped, purely poloidal magnetic field configuration. At the mid-plane, the BP field varies as $B_{\rm p}\propto r^{-1.25}$.

The 1D velocity dispersion of the fiducial $60\:M_\odot$ prestellar core, i.e., assuming virial equilibrium, is $1.09(M_c/60\:M_\odot)^{1/4}(\Sigma_{\rm cl}/1\:{\rm g\:cm}^{-2})^{1/4}\:{\rm km\:s^{-1}}$. 
In our simulation we adopt an isothermal equation of state with an effective sound speed, i.e., signal speed, of $c_{s} = 0.90\:{\rm km~s^{-1}}$. This choice is made so that the core is moderately sub-virial and will undergo gravitational contraction.

The gravitational field is treated with a simple approximation in which the mass of the star and the disk, residing outside of the simulation domain, are treated as a point mass. For the contribution of the potential of the envelope material, we assume a simple model of a fixed core envelope size, i.e., of radius $R_c$, and a fixed power law index describing the radial distribution, i.e., $\rho \propto r^{-3/2}$, but with the normalization of the profile adjusted to match the mass that is remaining in the envelope.

When the simulation starts, the core immediately begins to contract as the initial setup is unstable to gravitational collapse.
Initially, the plasma-$\beta$ (i.e., where $\beta\equiv P_{\rm gas}/P_{\rm mag}$) is slightly above unity in the core.
However, as the envelope collapses, the plasma-$\beta$ drops below unity, meaning that the magnetic field starts to dominate.
The collapse will therefore not be spherically-symmetric towards the protostar, but instead be guided along the field lines towards the mid-plane, notwithstanding the effects of the outflow and subsequent opening of a cavity; see \S\ref{sub:outflowstructure} and \S\ref{sub:cavityevolution}.

\subsection{Injection of the disk wind}
\label{sub:diskwindinjection}

We launch the disk wind through the injection region on the inner $x_1$ boundary, with $\dot{m}_{\rm inj}= \frac{1}{2}\frac{1}{10} \dot{m}_{*} = 0.05 \dot{m}_{*}$.
We also enforce that the injected outflow has the same momentum rate in the $x_1$ direction as in \citet{2014ApJ...788..166Z}.
Together, this can be used to constrain the injected density and velocity in the $x_1$ direction (perpendicular to the injection boundary).
As in Paper I, we then have an injected density:
\begin{equation}
  \rho_{\rm inj}=\begin{cases}
  \exp{(0.0289~r_{\rm cyl}/r_*)}\phi_{\rho} \rho_{\rm 0} & \quad r_{\rm cyl}<x_0\\
  2.77\bigg(\dfrac{r_{\rm cyl}}{x_0}\bigg)^{-1}\phi_{\rho} \rho_{\rm 0} & \quad r_{\rm cyl}\ge x_0\\
  \end{cases}
\label{rhoinjeq}
\end{equation}
and an injected $v_1$ velocity:
\begin{equation}
  v_{\rm inj}=(r_{\rm cyl}/r_*)^{-1/2}\phi_{\rm inj} v_{\rm K*},
\label{vinjeq}
\end{equation}
where $r_*$ is the stellar radius, $x_0=35.3r_*$, $r_{\rm cyl}$ is the distance from the $x_1$ axis, $\rho_{\rm 0}$ is the injection density at the axis, $v_{\rm K*}$ is the Keplerian speed on the stellar surface. 
$\phi_{\rho}$ and $\phi_{\rm inj}$ are time dependent dimensionless factors that are needed in order to obtain the desired mass flow and momentum rates of the inflowing wind, as is discussed in Paper I.

The velocity components of the injected flow in the 2- and 3-directions are set so that the flow is along the direction of the initial magnetic field lines. The injected flow is also given an additional toroidal velocity component:
\begin{equation}
v_{\rm \phi, inj}=0.23 \bigg(\frac{r_{\rm cyl}}{22.4 r_*}\bigg)^{-1/2}v_{\rm K*}.
\label{vphiinjeq}
\end{equation}

The values employed for $r_{\rm inj}$, $\rho_{\rm 0}$, $\dot{m}_{*}$, $\dot{m}_{\rm inj}$, and $\dot{p}_{\rm inj}$ are given for protostellar masses of $1$, $2$, $4$, $8$, $16$, and $24~{M_\odot}$ in Table~\ref{interptable}.

\begin{table*}
	\caption{Values of the radius of the injection region $r_{\rm i}$, the injected density along the axis $\rho_{\rm 0}$, the desired accretion rate $\dot{m}_{\rm acc}$, the desired injected mass flow rate $\dot{m}_{\rm inj}$, and the desired injected momentum rate $\dot{p}_{\rm inj}$ employed at the lower $x_1$ boundary for protostellar masses $m_*$. 
 }
	\begin{tabular}{cccccc}
           $m_*$ & $r_{\rm i}$ & $\rho_{\rm 0}$ & $\dot{m}_{\rm acc}$ & $\dot{m}_{\rm inj}$ & $\dot{p}_{\rm inj}$ \\
		${\rm [M_\odot]}$ & [au] & $[10^{-17}~{\rm g~cm^{-3}}]$ & $[10^{-4}{\rm M_\odot~yr^{-1}}]$ & $[10^{-5}{\rm M_\odot~yr^{-1}}]$ & $[10^{-3}{\rm M_\odot~km~s^{-1}~yr^{-1}}]$ \\
		\hline
		1 & 92 & $5.8$ & $1.0$ & $1.0$ & $5.9$ \\
		2 & 106 & $4.4$ & $1.4$ & $1.4$ & $9.9$ \\
		4 & 124 & $1.1$ & $2.0$ & $2.0$ & $9.4$ \\
		8 & 150 & $0.6$ & $2.7$ & $2.7$ & $14.2$ \\
		16 & 196 & $1.5$ & $3.2$ & $3.2$ & $41.2$ \\
		24 & 282 & $1.0$ & $3.3$ & $3.3$ & $49.5$ \\
	\end{tabular}
	\label{interptable}
\end{table*}

In the smoothing region, at $r_{\rm inj}<r<r_{\rm o}$, the velocity is gradually reduced by multiplying it by a factor $w={\rm cos}^2 [\frac{\pi}{2} (r-r_{\rm inj})/(r_{o}-r_{\rm inj})]$.
The initial density of the surrounding envelope is gradually joined with the density in the injection region by dividing the core density by $1+w (f_{\rm jump}-1)$, where $f_{\rm jump}$ is the ratio of the initial core density to the density in the injection region.

We note that, especially in the outflow cavity, if the density in a cell drops too low, the Alfv{\'e}n time step drops to such a low value that the simulation effectively grinds to a halt.
To avoid this, it is common practice in outflow simulations to implement a density floor, which prevents the Alfv{\'e}n time step from becoming extremely small.
However, including such a density floor means mass is being artificially added to the grid.
In this work, we have used a density floor that depends on height $x_1$ above the disk: $n_{\rm H,floor} = (x_1/10^5\:{\rm au})^{-1}\:{\rm cm^{-3}}$. The reason for this choice is that near the inner $x_1$ boundary where mass is accreting, we need a fairly large density floor to maintain a reasonable Alfv{\'e}n time step as the magnetic fields are stronger.
High above the disk, the density in the outflow cavity drops to values much below what the floor needs to be near the inner $x_1$ boundary, and hence the density floor in the outer part of the simulation box can be lower than in the inner part. 
We note that when mass is added to a cell in the simulation, we do not adjust the velocity of that cell, and as a consequence momentum is also added to the simulation. 

\section{Results}
\label{resultssection}

\subsection{Density, velocity and magnetic field structures}
\label{sub:outflowstructure}

\begin{figure*}
	\includegraphics[width=\textwidth]{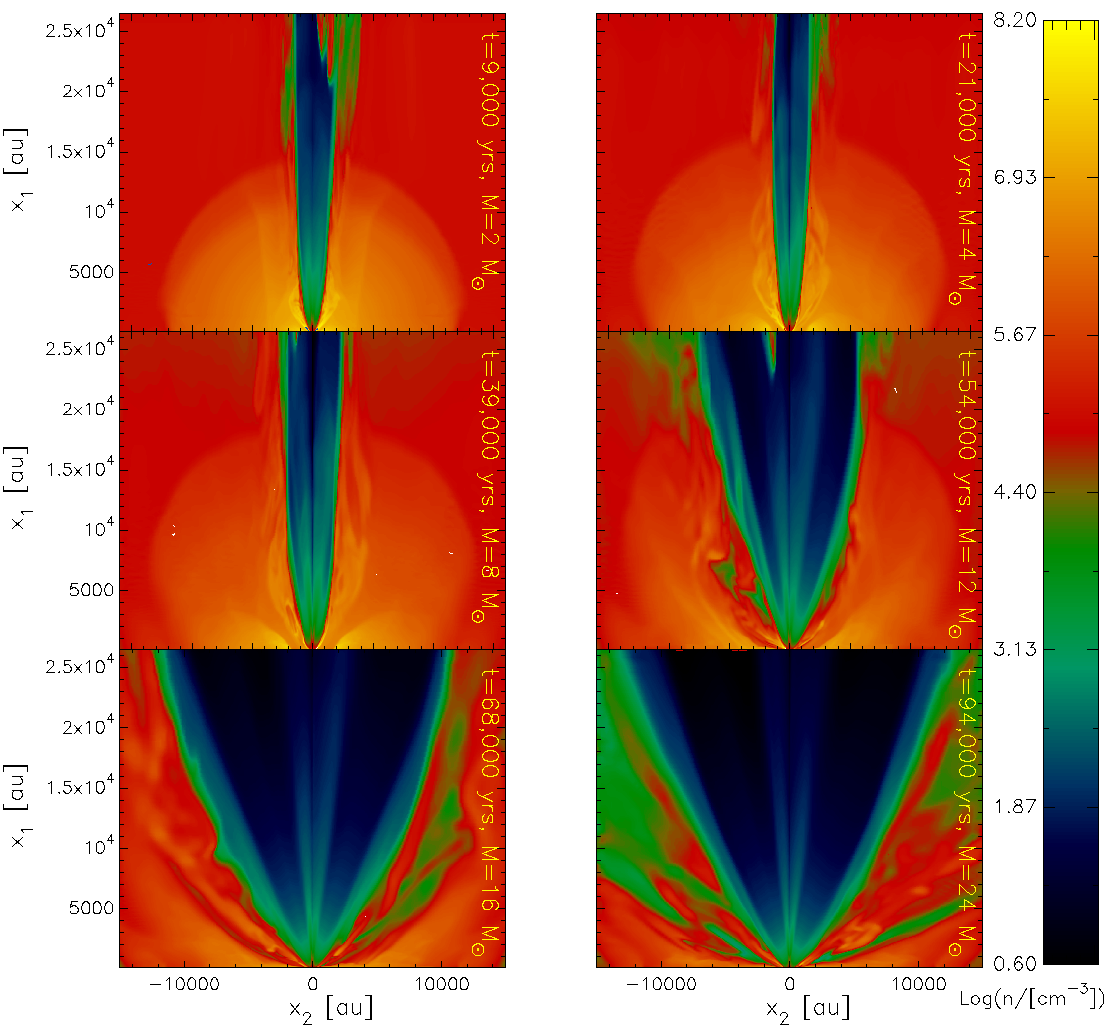}
	\caption{Slices of simulation results for density in the $x_1-x_2$ plane at $x_3 = 0$, with $x_1$ corresponding to the outflow axis. The top, middle and bottom rows show $m_*=2\:M_\odot$ and $4\:M_\odot$, $8\:M_\odot$ and $12\:M_\odot$, and $16\:M_\odot$ and $24\:M_\odot$, respectively. 
}
	\label{logdsummary}
\end{figure*}

\begin{figure*}
	\begin{interactive}{animation}{logdside.mpg}
		\includegraphics[width=0.9\textwidth]{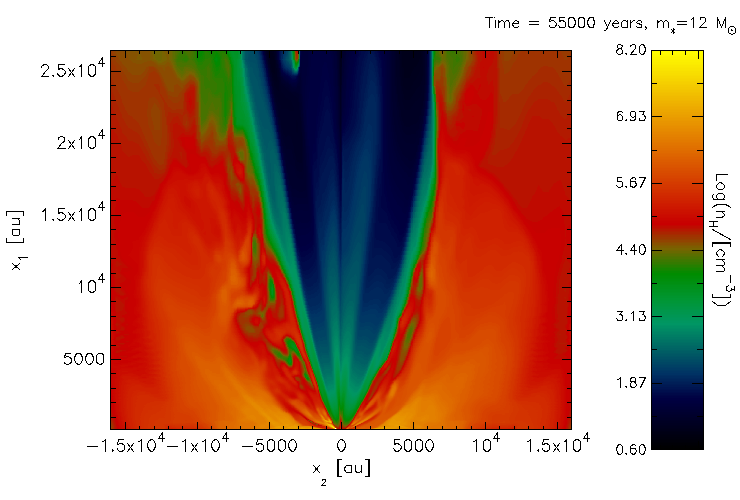}
	\end{interactive}
	\caption{Movie showing the temporal evolution of the number density in the $x_1-x_2$ plane at $x_3=0$, i.e., the same as the individual time slices shown in Fig.~\ref{logdsummary}. The movie runs for $\sim$19 seconds and shows the evolution of the outflow cavity over a period of 97,000 yr. In the top right corner, the simulation time in yr and the current protostellar mass are indicated.
 }
	\label{logdevol}
\end{figure*}

\begin{figure*}
	\includegraphics[width=\textwidth]{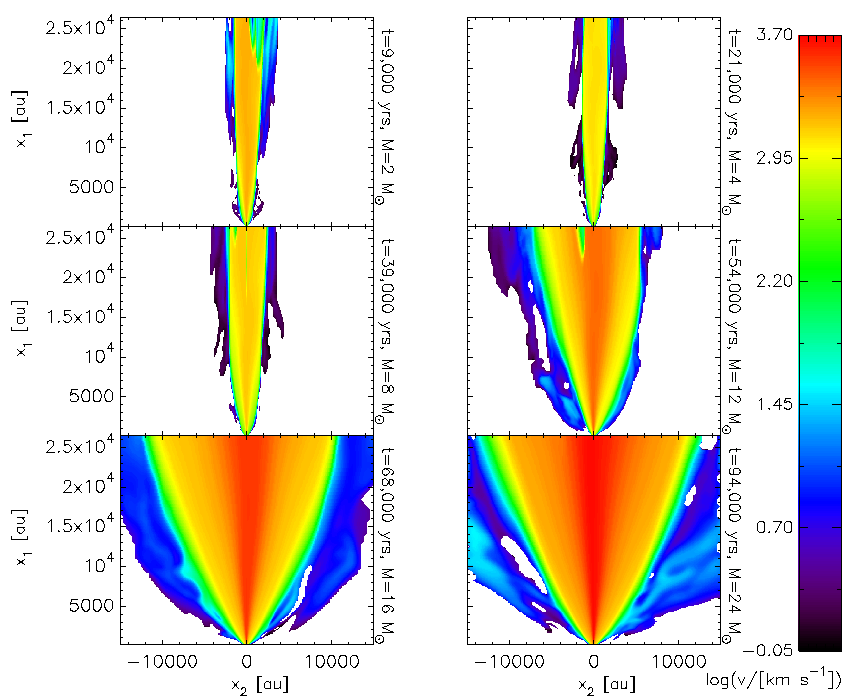}
	\caption{Slices in the $x_1-x_2$ plane at $x_3 = 0$ of simulation results for total velocity, $v$, but only showing cells with $v_1>0.9\:{\rm km\:s}^{-1}$ to highlight outflowing gas. The top, middle and bottom rows show $m_*=2\:M_\odot$ and $4\:M_\odot$, $8\:M_\odot$ and $12\:M_\odot$, and $16\:M_\odot$ and $24\:M_\odot$, respectively.}
	\label{logvsummary}
\end{figure*}

\begin{figure*}
	\includegraphics[width=\textwidth]{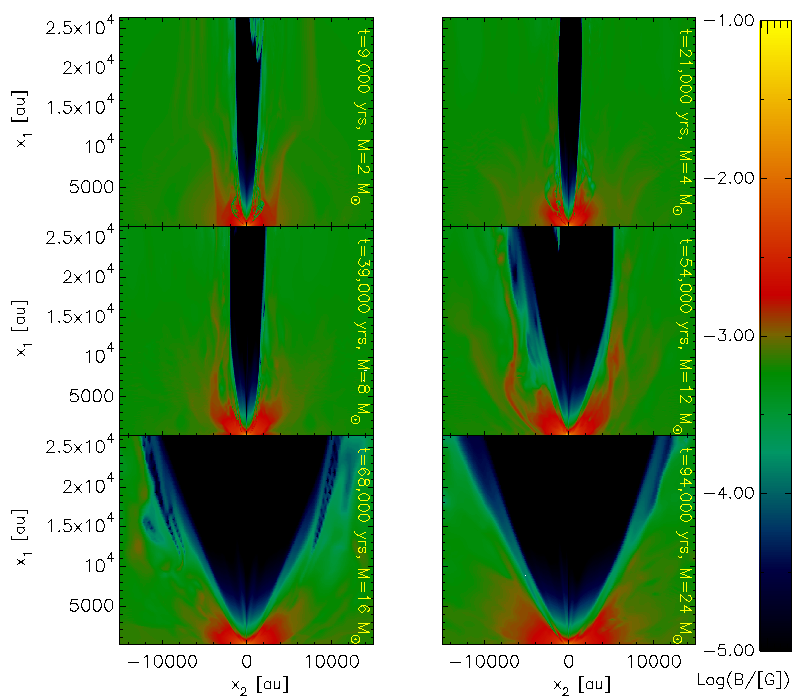}
	\caption{Slices of simulation results for magnetic field strength, $B$, in the $x_1-x_2$ plane at $x_3 = 0$. The top, middle and bottom rows show $m_*=2\:M_\odot$ and $4\:M_\odot$, $8\:M_\odot$ and $12\:M_\odot$, and $16\:M_\odot$ and $24\:M_\odot$, respectively.}
	\label{logbsummary}
\end{figure*}

We have simulated the evolution of the protostellar core for $\sim10^5\:$yr as the protostar grows from $m_*=1\:{M_\odot}$ to about $26\:{M_\odot}$. In Fig.~\ref{logdsummary} we show slices of the density structure in the $x_1-x_2$ plane at $x_3=0$. These images show the general structure of the disk-wind outflow cavity as it gradually carves open a larger and larger volume from the initial core infall envelope. Concurrent with this evolution of the outflow cavity, we also see the collapse of the infall envelope down towards the central midplane base of the core. A movie showing the evolution of this structure is shown in Fig.~\ref{logdevol}. During the course of the evolution the range of densities present in the simulation extends from $n_{\rm H}\sim 4\:{\rm cm}^{-3}$ (in the outflow cavity) to $\gtrsim 10^8\:{\rm cm}^{-3}$ (in the inner infall envelope). 

Figure~\ref{logvsummary} shows the magnitude of the outflowing velocity along the $x_1$ direction, i.e., $v_1>0.9\:{\rm km\:s}^{-1}$, for the same slices through the simulation domain shown in  Fig.~\ref{logdsummary}.
At any given evolutionary stage, the highest velocities are found close to the central axis of the outflow cavity. At the earliest stages shown in Fig.~\ref{logvsummary}, i.e., $m_*=2\:M_\odot$, these velocities are already $\sim 2,000\:{\rm km\:s}^{-1}$. By the later stages with $m_*=24\:M_\odot$, these velocities have risen to $\sim 5,000~{\rm km\:s^{-1}}$.

Figure~\ref{logbsummary} shows the magnitude of the total magnetic field strength for the same slices through the simulation domain shown in  Fig.~\ref{logdsummary}. The largest magnetic field strengths are $\sim 100\:{\rm mG}$ near the base of the outflow and inner infall envelope.
In the outflow cavity, the magnetic field strength is much lower than in the infall envelope, with values at low as $\sim 0.01\:$mG.

\subsection{Evolution of the outflow cavity opening angle}
\label{sub:cavityevolution}

\begin{figure*}
	\includegraphics[width=0.95\textwidth]{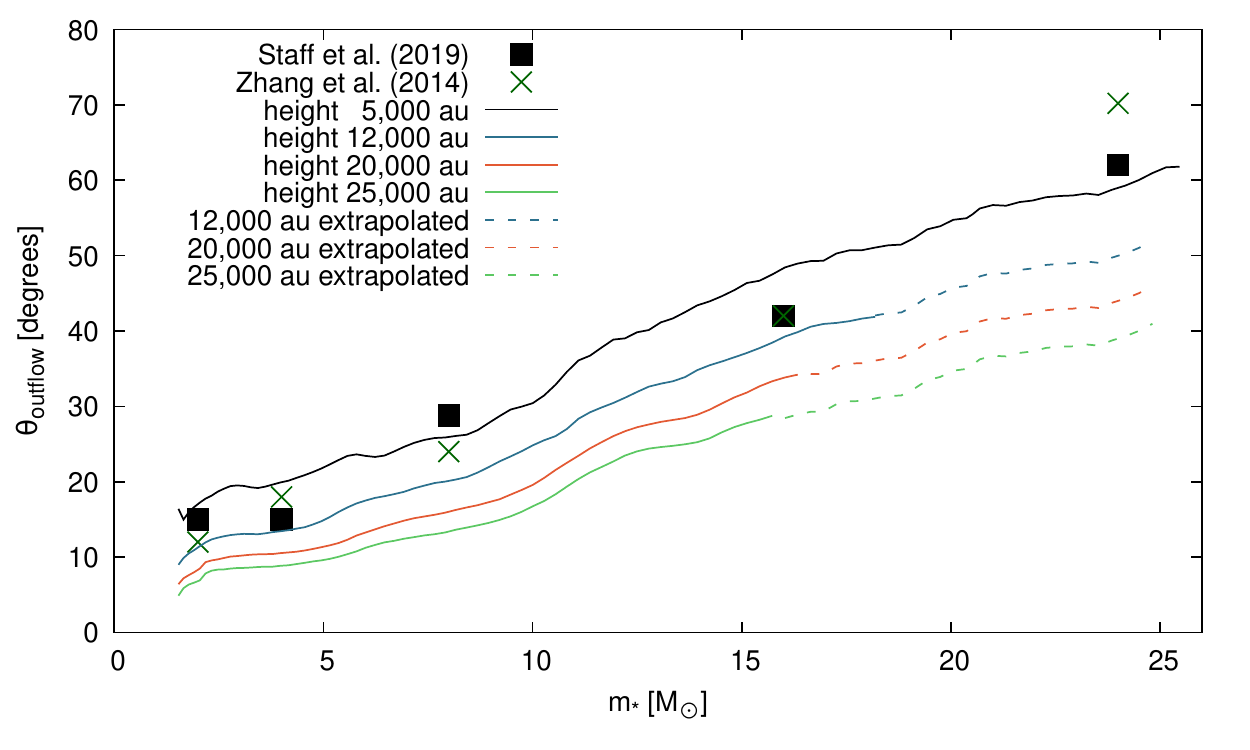}
	\caption{Outflow cavity opening angle measured at different heights above the disk (solid lines). Extrapolated estimates (dashed lines) are needed once the cavity nears the simulation boundary at a given height (see text).
  Also shown are the outflow cavity opening angles found in the numerical models of Paper I (without pre-clearing; squares) and the semi-analytic models of \citet{2014ApJ...788..166Z} (crosses).
  }
	\label{openingangleplot}
\end{figure*}

To evaluate the opening angle of the outflow cavity at a given height $x_1$, we first calculate the area $A$ in the $x_2$-$x_3$ plane of the outflowing matter that has $v_1>0.9~{\rm km\:s^{-1}}$. 
We then approximate the outflow as having a conical shape with a circular cross section of area $A=\pi r^2$,  giving $r=\sqrt{A/\pi}$, and then find the half opening angle of that cone, $\tan(\theta_{\rm outflow}) = r/x_1$.
In Fig.~\ref{openingangleplot} we show the evolution of the calculated opening angle over time for several different heights above the disk. These direct estimates of the opening angles are stopped when the outflow cavity region approaches the lateral edges of the simulation domain. Beyond this point, shown with dashed lines, we make an approximate estimate for opening angle at a given height via linear extrapolation from the closest lower height where the geometry of the outflow is still contained within the domain. Given the apparent direction of curvature of the outflow cavity towards the outflow axis, e.g., see Fig.~\ref{logvsummary} for $m_*=16\:M_\odot$, we expect that this method will tend to overestimate the opening angle.

From our results we see that the outflow cavity opening angle is larger at lower heights (e.g., at 5,000 au), and is smaller at larger heights due to collimation of the outflow. In other words, the outflow cavity is not truly conical (as is evidenced in Figs.~\ref{logdsummary} and \ref{logdevol}). Considering a fidcuial height equal to the initial radius of the core, i.e., 12,000~au, we see that the outflow cavity opening angle has achieved a value of about $10^\circ$ at the earliest stages of the simulation, i.e., when $m_*=2\:M_\odot$. It then rises slowly until $m_*\sim 4\:M_\odot$. After this it increases at a slightly faster rate, reach about $42^\circ$ by the time $m_*=18\:M_\odot$, i.e., the last stage where it can be directly evaluated in the simulation domain. An extrapolation based estimate at $m_*=24\:M_\odot$ yields $\theta_{\rm outflow}\simeq 50^\circ$.

In Figure~\ref{openingangleplot} we also compare our results to those of Paper I (without pre-clearing), which were calculated at the top of the grid in those simulations, i.e., at a height of about $12,000$ au. We note that in Paper I, with models run at fixed $m_*$, it was somewhat uncertain at which time to evaluate the results for the opening angle. Paper I also considered a case ``with pre-clearing'' that attempted to allow for the earlier stages of evolution and these yielded larger opening angles at the later stages, i.e., about $50^\circ$ at $m_*=16\:M_\odot$ and $78^\circ$ at $24\:M_\odot$. We find that our new simulations with a continuous evolution followed from low to high values of $m_*$ yield moderately smaller cavity opening angles than the results of Paper I, with the biggest differences being at the highest masses.

We also compare our results to the opening angles predicted by the semi-analytic model of \citet{2014ApJ...788..166Z}, following the method of \citet{2000ApJ...545..364M}, which is based on the condition of whether the material in a given direction can be accelerated to the escape speed. We find that our numerical results predict a moderately narrower outflow cavity geometry than this semi-analytic model, with the difference being about $20^\circ$ by the end of the simulation.

\subsection{Mass and momentum fluxes of the outflow}
\label{sub:outflowfluxes}

\begin{figure}
        \includegraphics[width=0.45\textwidth,trim=0 20 0 0]{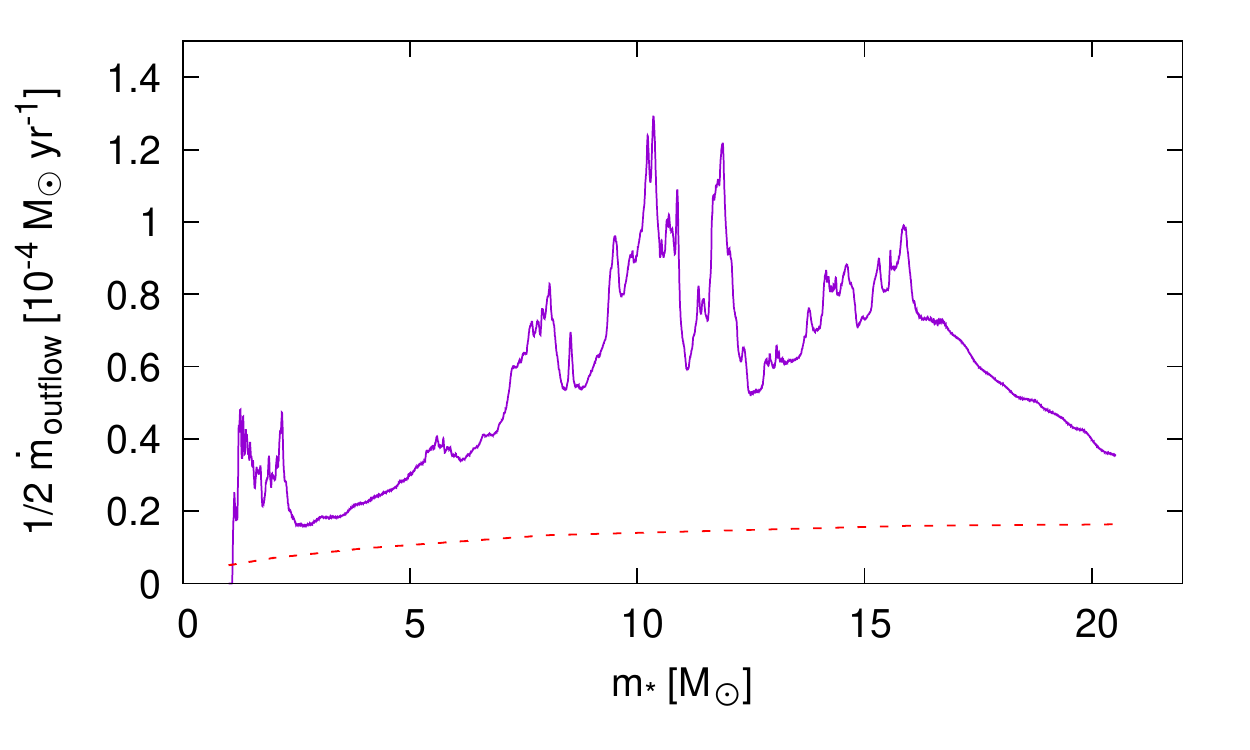}
        \includegraphics[width=0.45\textwidth]{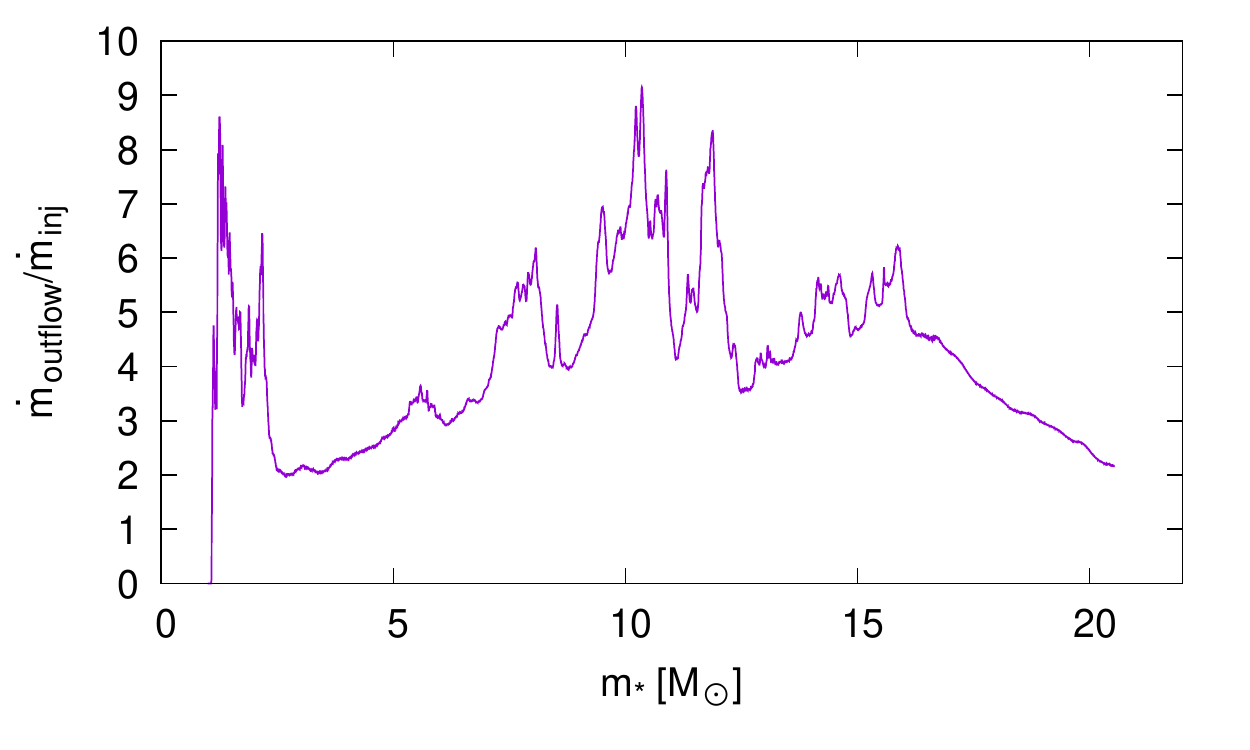}
        \includegraphics[width=0.45\textwidth,trim=0 20 0 0]{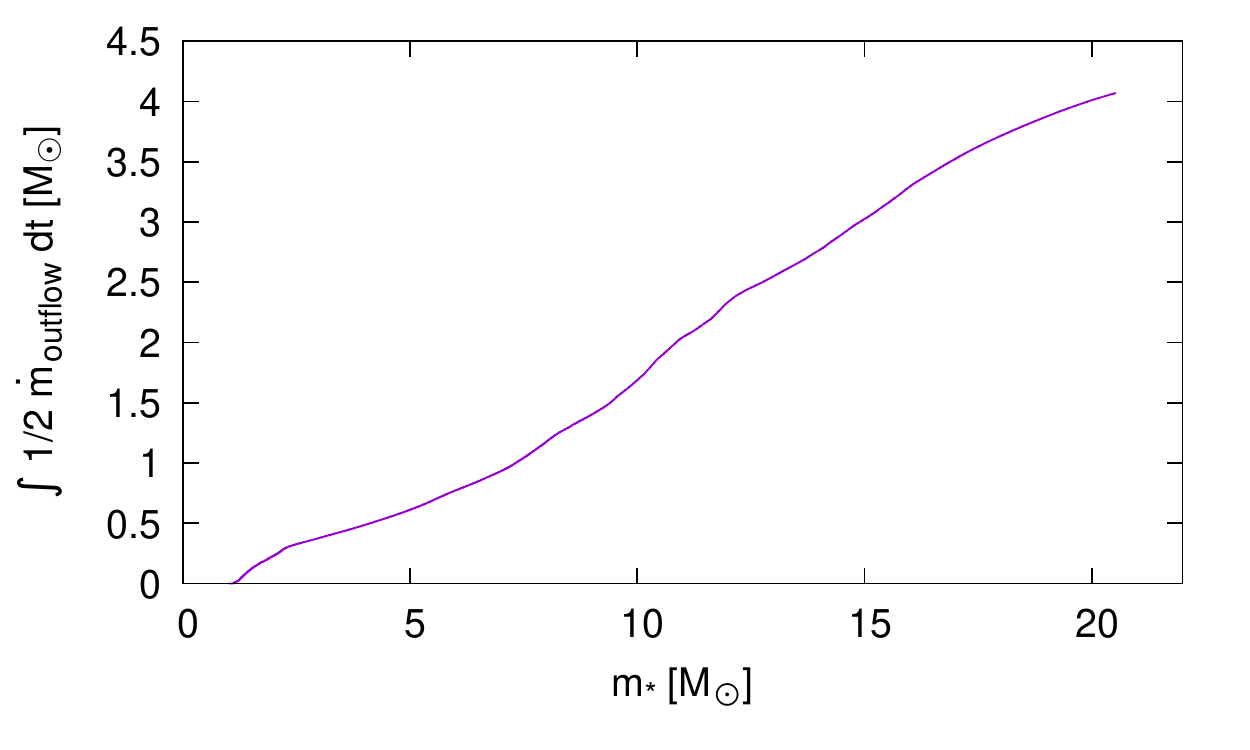}
	\caption{{\it (a) Top:} Evolution of outflow mass flux through the top of the simulation domain ($x_2-x_3$ face at $x_1=25,000\:$au) (purple solid line). 
 The red dashed line shows the injected mass flow rate of the outflow. 
 {\it (b) Middle:} Ratio of the mass flow rate out of the top of the simulation box to the injected mass flow rate at base of the outflow.
  {\it (c) Bottom:} Evolution of total mass that has left the top of the simulation domain by being swept-up by the outflow.
 }
	\label{mflux}
\end{figure}

\begin{figure}
        \includegraphics[width=0.45\textwidth,trim=0 20 0 0]{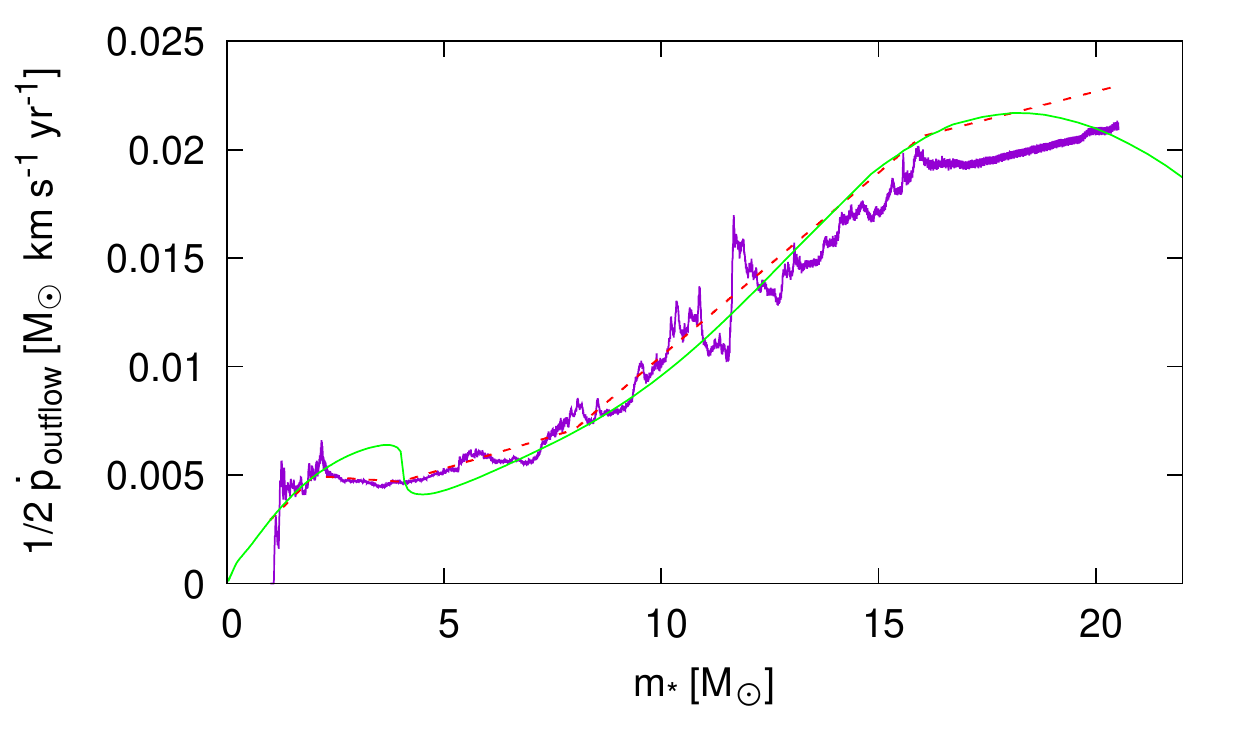}
        \includegraphics[width=0.45\textwidth]{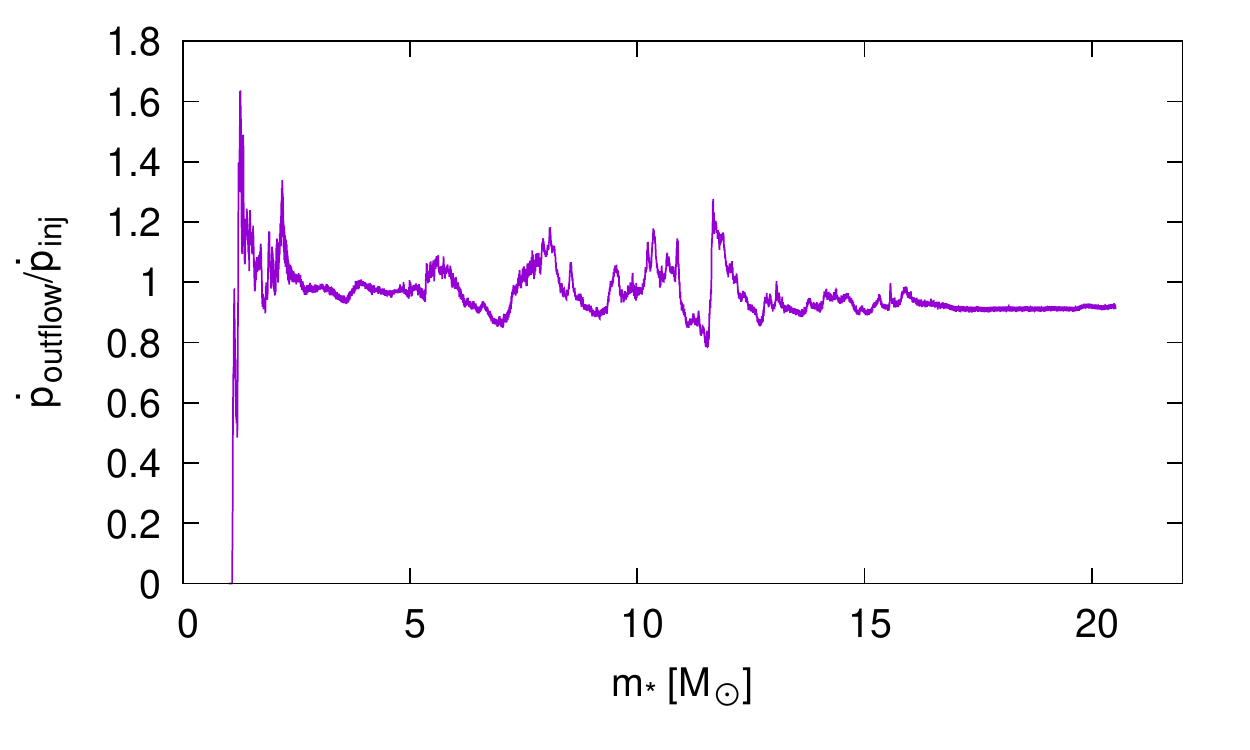}
        \includegraphics[width=0.45\textwidth,trim=0 20 0 0]{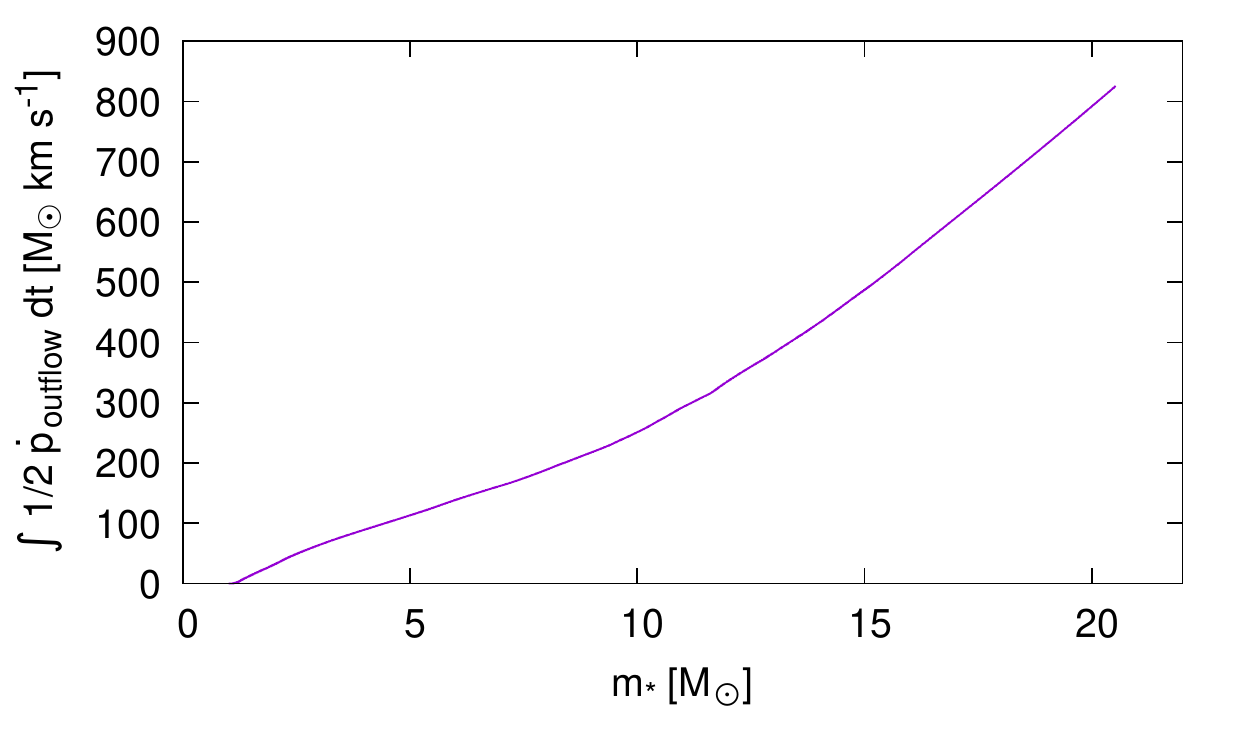}
	\caption{{\it (a) Top:} Evolution of outflow momentum flux through the top of the simulation domain ($x_2-x_3$ face at $x_1=25,000\:$au) (purple solid line). The red dashed  line shows the injected momentum flux at the base of the outflow. The green solid line shows the momentum flux injected in the semi-analytic model of \citet{2014ApJ...788..166Z}.
 {\it (b) Middle:} Evolution of the ratio of the momentum flux through the top of the simulation domain to the injected momentum flux at the base of the outflow.
  {\it (c) Bottom:} Evolution of the total momentum that has left the top of the simulation domain.
    }
	\label{momflux}
\end{figure}

We evaluate the rate at which mass flows out of the top of the simulation box at the $x_2-x_3$ boundary face via $(1/2)\dot{m}_{\rm outflow}= \int \rho v_1 dA$, i.e., performing the summation over the actual area of the outflow with no assumption of it being circular and equating this to half the total mass flux in a bipolar protostellar outflow. The evolution of this outflowing mass flux is shown in Figure~\ref{mflux}a.

Initially, there is a transient phase with a fairly high mass flux out of the simulation box of $\sim4\times10^{-5}\:M_\odot\:{\rm yr}^{-1}$ while the outflow cavity is being cleared out.
After this the mass flow rate grows from about $2\times10^{-5}\:M_\odot\:{\rm yr}^{-1}$ to $\sim 1\times10^{-4}\:M_\odot\:{\rm yr}^{-1}$ by the time the star has reached $\sim 10\:{M_\odot}$. We note that the mass flux exhibits moderate, $\sim30\%$, fluctuations during this evolution. After this the mass flux stops increasing and exhibits more dramatic fluctuations during the evolution to $m_*=16\:M_\odot$. After this, it shows a more steady, smooth decline, which is mostly caused by the outflow cavity expanding beyond the size of the top face of the simulation domain. For this reason, we do not calculate the mass flow rate out of the grid for masses beyond $\sim 20\:M_\odot$: i.e., at this stage a significant amount of mass is now leaving across the side boundaries (as can be observed in the movie in Fig.~\ref{logdevol} and in Fig.~\ref{logvsummary}).

Figure~\ref{mflux}b shows the ratio of the mass flux leaving the top of the simulation domain to the mass injected at the base of the outflow. After the initial peak associated with first breakout of the outflow, this ratio is about 2, but then rises up to a peak just below 10 when $m_*=10\:M_\odot$. At higher masses it generally declines, but with large fluctuations, eventually reaching values near 2 again.

Figure~\ref{mflux}c shows the time evolution of the total mass that has left the top of the simulation domain. We find that more than $4\:M_\odot$ has left the grid as part of the outflow by the time the protostar reaches $20\:M_\odot$.

Figure~\ref{momflux}a shows the momentum flux passing through the top of the simulation domain, evaluated as $\dot{p}=\int \rho v_1^2 dA$. As in Fig.~\ref{mflux}, we cut off the measurements when substantial mass and momentum start to leave the domain through the side boundaries.
We find that the momentum flux leaving the domain stays approximately constant at about $0.005\:M_\odot\:{\rm km\:s^{-1}\:yr^{-1}}$, until the star reaches $\sim7\:M_\odot$.
Then it increases to reach nearly $0.02\:M_\odot\:{\rm km\:s^{-1}\:yr^{-1}}$ when the star is $\sim16\:M_\odot$. It then continues to increase, but at a slower rate. However, at this stage we begin to lose track of mass that is leaving through the sides of the domain.

Figure~\ref{momflux}a also shows the injected momentum flux at the base of the outflow.
In general, as expected, we see a very good agreement between the injected and ejected momentum fluxes, with the largest deviation occurring at late times due to some outflow material leaving via the sides of the domain. The ratio of these momentum fluxes is shown explicitly in Figure~\ref{momflux}b.

Figure~\ref{momflux}c shows the total momentum that has left via the top of the simulation domain. This grows steadily to reach $\sim 800\:M_\odot\:{\rm km~s^{-1}}$ by the time the protostar has reached $\sim20\:M_\odot$.

\subsection{Star formation efficiency}
\label{sub:sfe}

Here we evaluate the star formation efficiency (SFE), i.e., the ratio of the final stellar mass to the initial core mass, that is implied by our simulation results. After 100,000 years, the protostar has grown to $m_*\simeq 26\:M_\odot$. Thus we estimate that $\bar{\epsilon}_{*f}\geq 0.43$. This is a lower limit since in our model the disk has a mass of $m_{\rm disk} = (1/3) m_* \simeq 9\:M_\odot$ and a significant portion of this material is expected to be able to accrete to the star. If the only process diverting material from the accretion disk is injection into the disk wind with $\dot{m}_w=0.1\dot{m}_*$, then the final stellar mass would be at least $34\:M_\odot$, i.e., $\bar{\epsilon}_{*f}\geq 0.56$. It is possible that a larger fraction of material could be diverted from the accretion disk if other forms of feedback, especially disk photoevaporation, are significant. However, \citet{2017ApJ...835...32T} considered such models and found that disk photoevaporation was relatively unimportant compared to the disk wind mass flux for this mass and accretion rate regime.

The above estimates are likely to still be lower limits, since there is still $12\:M_\odot$ ($3\:M_\odot$ from the initial core and $9\:M_\odot$ from the surrounding clump) remaining in the simulation domain, i.e., $24\:M_\odot$ in the global, mirrored domain. 
One expects that a significant fraction of this material would be accreted to the central protostar. In the case that all of the remaining initial core mass is accreted, i.e., $6\:M_\odot$, then this would thus result in a SFE of $\bar{\epsilon}_{*f}\simeq 0.67$.

Comparing the semi-analytic model of \citet{2014ApJ...788..166Z}, they also reached a final value of $m_*=26\:M_\odot$. Thus, with the same considerations of residual disk accretion, they expect to reach $\bar{\epsilon}_{*f}\geq 0.56$. However, their model at this point would be exhausted of gas and so this would be the final estimate of SFE. Thus we conclude that the expected SFE from our numerical model is moderately ($\sim20\%$) larger than that predicted by the semi-analytic model. This is consistent with the generally smaller outflow opening angles found during the course of the evolution in the numerical model compared the \citet{2014ApJ...788..166Z} semi-analytic model (see Fig.~\ref{openingangleplot}). 

However, we note that in the fiducial TCA model of \cite{2003ApJ...585..850M}, the initial core is expected to interact with significant surrounding clump gas during its collapse to a protostar, so with this consideration the results of \citet{2014ApJ...788..166Z} for the final stellar mass, $m_{*f}$, are also lower limits. If SFE is defined with respect to the initial core mass, then the values of $\bar{\epsilon}_{*f}$ would also be lower limits.

\subsection{Outflow mass spectra}

\begin{figure*}
\includegraphics[width=0.9\textwidth]{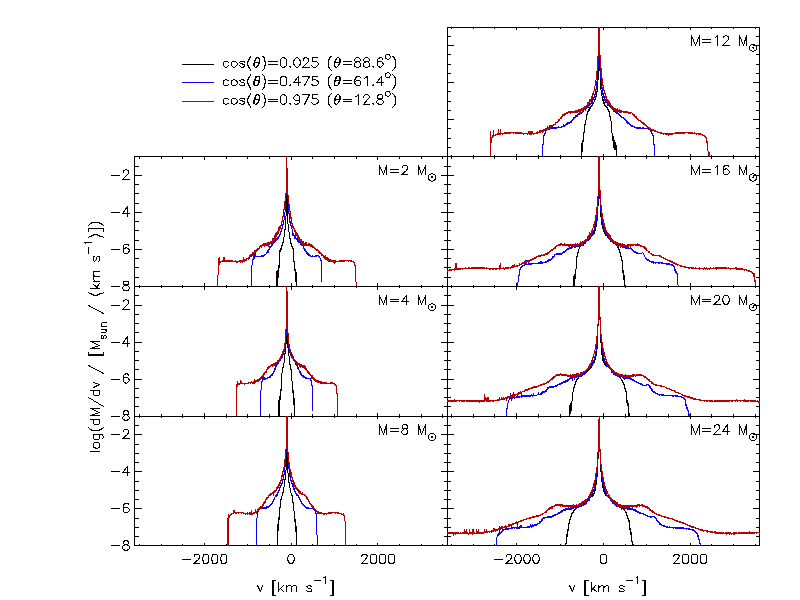}
\caption{Distribution of outflow mass with line of sight velocity for material within a global (i.e., mirrored) simulation domain at various evolutionary stages (i.e., protostellar masses) and as viewed at different inclination angles, $\theta_{\rm view}=12.8^\circ, 61.4^\circ, 88.6^\circ$. 
}
\label{masshistmulti}
\end{figure*}

One method of comparing our model results with observed systems is via the distribution of outflowing gas mass with line of sight velocity velocity, i.e., ``mass spectra'', since this can be inferred from observations of CO emission lines. Note, in this paper we will not make synthetic CO spectra of our models, deferring this step to a future work. To produce the distribution of mass with line of sight velocity, we need to produce a ``global'' simulation domain, which is achieved by mirroring our simulation grid about the $x_1=0$ boundary, i.e., the disk plane. In this way we produce a symmetric bipolar outflow structure, which we then view at various angles, $\theta_{\rm view}$, to the outflow axis. Note, $\theta_{\rm view}=0^\circ$ is defined as a line of sight that is parallel to the outflow axis.

Figure~\ref{masshistmulti} shows the mass spectra within the global domain at various evolutionary stages. Note, these spectra include all gas, i.e., both outflowing and infalling material. We have chosen three values of $\theta_{\rm view}$ that are part of the grid of uniformly sampled grid of ${\rm cos}\:\theta_{\rm view}$ values in the radiative transfer models of \citet{2018ApJ...853...18Z}. The mass spectra show a sharp peak at low velocities, and, except for $\theta_{\rm view}$ values close to $90^\circ$, long tails to larger velocities.
As the protostellar mass increases, we find more mass at larger velocities. For $m_*>16\: M_\odot$, the largest velocities are $>3000\:{\rm km~s^{-1}}$ when the system is viewed close to the outflow axis. One point to note is that between $2\:M_\odot$ and $4\:M_\odot$, the maximum velocities decrease somewhat. This is due to the protostellar radius (which also sets the inner disk radius) growing from $3.45\:R_\odot$ at $2\:M_\odot$ to $20.5\:R_\odot$ at $4\:M_\odot$. The injection velocity of the outflow is proportional to the Keplerian speed at the launching point ($v_{\rm Kep}\propto m_*^{1/2}~r^{-1/2}$; Eq.\ \ref{vinjeq}). Hence, the highest velocity outflow is launched from the inner disk and, as the inner disk radius expands, the velocity of the material launched from the inner disc decreases, even though the central mass is growing. We use these mass spectra in the next subsection to make detailed comparisons to some observed massive protostars.

\subsection{Comparison with observed outflow mass spectra}
\label{sub:compare_obs}

\begin{figure*}
\includegraphics[width=0.9\textwidth]{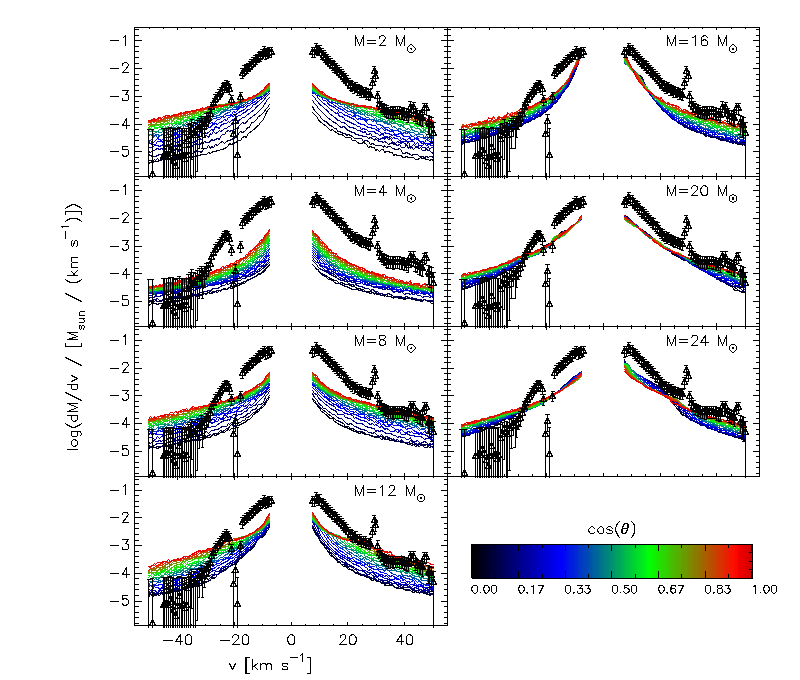}
\caption{The mass velocity spectra from the simulation compared to that from observations of G35.20-0.74N \citep{2022ApJ...936...68Z} for velocities less than $\pm 50~{\rm km~s^{-1}}$.}
\label{masshistg35}
\end{figure*}

\begin{figure*}
\includegraphics[width=0.9\textwidth]{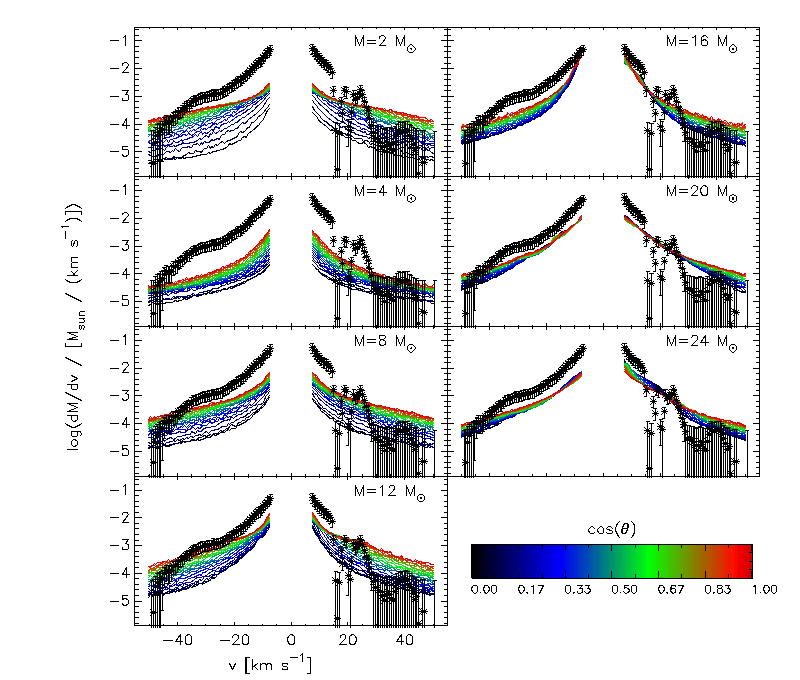}
\caption{The mass velocity spectrum from the simulation compared to that from observations of G339.88-1.25 \citep{2019ApJ...873...73Z}, for velocities less than $\pm 50~{\rm km~s^{-1}}$.}
\label{masshistg339}
\end{figure*}

\begin{figure}[ht]
\includegraphics[width=0.99\columnwidth,trim=40 0 10 0 ]{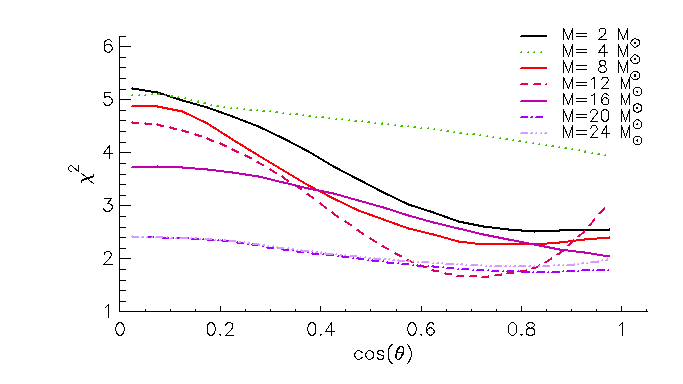}
\includegraphics[width=0.99\columnwidth,trim=40 0 10 0]{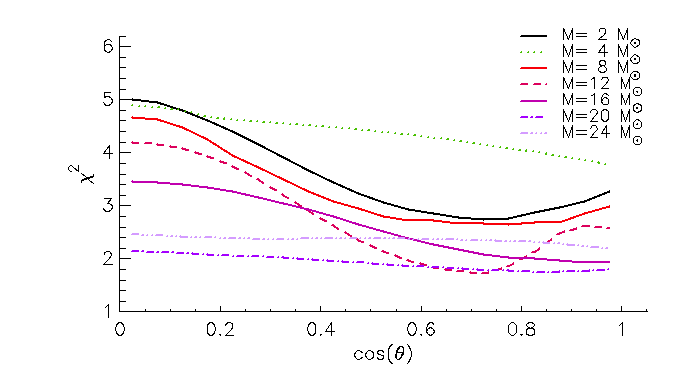}
\caption{Dependence of $\chi^2$ derived from fitting our simulated mass spectra for different evolutionary stages (i.e., various values of $m_*$) to the observational data of massive protostars G35.2 (top) and G339 (bottom) as a function of the cosine of the viewing angle.
}
\label{chicos}
\end{figure}

In Figures~\ref{masshistg35} and~\ref{masshistg339} we compare the simulation outflow mass spectra to equivalent outflow mass spectra of G35.20-0.74N and G339.88-1.26 (hereafter G35.2 and G339) as derived from ALMA observations of CO(2-1) line emission by \citet{2022ApJ...936...68Z} and \citet{2019ApJ...873...73Z}, respectively. Note, the observed line emission from these sources was extracted from regions of $\sim$25,000~au in radial size centered on the protostars, similar to the size of our simulation box. We consider a velocity range of $\pm 50\:{\rm km\:s^{-1}}$ and exclude the inner $\pm 10\:{\rm km\:s^{-1}}$, which is affected by the presence of ambient clump gas.

To quantify the differences between the models and observations, we calculate the reduced $\chi^2$ between the two, following the method of \citet{2018ApJ...853...18Z} (developed for spectral energy distribution fitting), as:
\begin{equation}
  \chi^2 = \frac{1}{N} \sum_{i}\left[\frac{ m_{\rm i,data}- m_{\rm i,sim}}{\sigma}\right]^2,
  \label{eq:chisq}
\end{equation}
where $N$ is the number of data points, $m_{\rm i,data}$ and $m_{\rm i,sim}$ are the mass in the i'th velocity bin in the observed data and in the simulation, and $\sigma$ is the uncertainty on the observed data. 
The uncertainty in the data is assumed to be comprised of a systematic uncertainty of $40\%$ and a noise level that is $\sim6\times10^{-5}~{\rm M_\odot/(km~s^{-1})}$ (for both G35.2 and G339). Note that while the mass spectra are shown in log space, we perform the $\chi^2$ fitting in linear space.

As seen in Figure~\ref{masshistg35}, G35.2's outflow mass spectrum at negative velocities is affected by a significant absorption feature at $-20\:{\rm km\:s^{-1}}$, which may be due to other molecular cloud components along the line of sight. Thus, for this source we restrict fitting to only the positive velocity range. Figure~\ref{masshistg339} shows that G339's mass spectrum at positive velocities is similarly affected by absorption features and so here we only fit to the negative velocity range.

Each of the panels in Figures~\ref{masshistg35} and~\ref{masshistg339} shows the models at a particular evolutionary stage as seen over the full range of viewing angles, i.e., uniformly sampling ${\rm cos}\:\theta_{\rm view}$ from 0.025 to 0.975 in steps of 0.05. We can see that at small values of $m_*$ the models generally fail to to match the observational data. In particular, they underpredict the amount of outflowing gas at low and intermediate velocities. For G35.2, there is a better agreement in the shape of the mass spectrum when $m_*\sim 16\:M_\odot$ to $24\:M_\odot$, although the model is systematically low by a factor of about 3. For G339, the shape of the mass spectrum has a best match when $m_*\sim 20\:M_\odot$, but is again low be about a factor of 3. We note that such systematic offsets could be explained, at least in part, by uncertainties in the conversion of CO(2-1) line flux to mass. The difference could also simply be due to the observed systems being more massive protostellar cores, i.e., involving an initial core mass that is $>60\:M_\odot$. Within the context of the Turbulent Core Accretion model, there is also the additional parameter of $\Sigma_{\rm cl}$, which could be varied from the fiducial value of $1\:{\rm g\:cm}^{-2}$ assumed here. 

Given the above considerations, we do not attempt to adjust our models further to find a better match to the data, since such a step will likely require running a much larger grid of simulations to explore the $M_c$ and $\Sigma_{\rm cl}$ parameter space. Nevertheless, with the context of the models we have presented, there is formally a best fitting model for each of G35.2 and G339. To illustrate these and the dependence of $\chi^2$ on model parameters, in Figure~\ref{chicos} we plot $\chi^2$ versus ${\rm cos}\:\theta_{\rm view}$ for all the considered models at various evolutionary stages.
Again, we can see that the observations are more consistent with higher protostellar masses.
However, in these higher mass cases, we note that the goodness of fit does not depend very sensitively on the viewing angle.

\subsection{Comparison to other observational metrics of massive protostars}

The mass flow rate out of the simulation box (see Fig.~\ref{mflux}) starts out at a few $\times 10^{-5}\:M_\odot\:{\rm yr^{-1}}$ for the first $\sim50,000$ years until the star reaches $\sim10\:M_\odot$, before increasing to more than $10^{-4}\:M_\odot\:{\rm yr}^{-1}$ and becoming quite variable during the latter parts of the simulation. 
The momentum flux out of the simulation box (Fig.~\ref{momflux}) is, meanwhile, about $5\times10^{-3}\:M_\odot\:{\rm km\:s^{-1}\:yr^{-1}}$ for the first $\sim40,000$ years until the star reaches $\sim8\:M_\odot$, after which the momentum rate grows steadily to $\sim2\times10^{-2}\:M_\odot\:{\rm km\:s^{-1}\:yr^{-1}}$, and also shows time-variable behaviour.
Such values are in general agreement with observations of outflows from massive protostars \citep{2004A&A...426..503W,2015MNRAS.453..645M,2019NatCo..10.3630F}, although it should be noted that there are significant uncertainties associated with the observational derivation of these mass and momentum fluxes.

There have been a few measurements of magnetic field strengths in the outflows of massive protostars. In Orion Source I, which is thought to be $10-20\:M_\odot$ protostar \citep[e.g., see discussion in][]{2020ApJ...896..157H}, the magnetic field strength was estimated to be 30 mG on a scale of a few hundred au. This is in reasonable agreement with our simulations on similar scales (Fig.~\ref{logbsummary}).

\section{Discussion}
\label{discussionsection}

\subsection{Comparison with previous simulation studies}
\label{sub:comparesims}

Here we discuss how our simulation results compare to those of other relevant studies of massive star formation, mostly restricting our consideration to those including protostellar outflow feedback with magnetohydrodynamic (MHD) simulations. The simulation we have presented, in addition to its initial core, has a well defined boundary condition during the evolution for the input protostellar outflow, which is tied to the evolution of the fiducial massive protostar in the Turbulent Core Accretion model \citep{2003ApJ...585..850M,2014ApJ...788..166Z}. One comparable non-MHD simulation is that of \citet{2018A&A...616A.101K}, who presented a simulation of a massive protostar forming from a surrounding mass reservoir from $100\:M_\odot$ to $1000\:M_\odot$. The simulation code {\it Pluto} \citep{2012ApJS..198....7M} was utilized with a logarithmically spaced spherical coordinate grid assuming axial and midplane symmetry of the system. Feedback from radiation pressure, ionization and injected protostellar outflows was included. However, the simulation did not include magnetic fields. With this caveat in mind, their most comparable case involved a 0.1~pc radius core with mass of 100$\:M_\odot$, which formed a star with mass of $66\:M_\odot$ when only protostellar outflows were included, dropping to a minumum value of $51\:M_\odot$ when outflows and radiation pressure were both implemented.

In contrast, the following simulation studies generally present collapse of a fully 3D gas structure to a sink particle representing a protostellar source. For example, \citet{2020AJ....160...78R} performed radiation MHD simulations of a collapsing $150\:{M_\odot}$ core (significantly more massive than the $60\:M_\odot$ core we consider in this study), and followed the evolution until the star reached a mass of $33.64\:{M_\odot}$.
They found that once the stellar mass reached about $30\:{M_\odot}$, radiation pressure created by the central star starts driving an expanding bubble.
Radiative effects like this could potentially be relevant in our case if we continued the simulation beyond $30\:{M_\odot}$ \citep[see also][]{2017ApJ...835...32T}.

\citet{2021arXiv210910580C} compared collapse simulations of a $100\:{M_\odot}$ core in several scenarios: without magnetic fields, with ideal MHD, and with ambipolar diffusion.
In the case of the non-magnetized simulation, they found a very weak outflow dominated by episodes of accretion bursts. In their ideal MHD simulation, they found that an increased pressure in the central region, due to increased stellar luminosity and build-up of magnetic field, causes the outflow to almost disappear when the protostar reaches $\sim 10\:{M_\odot}$. However, this behaviour is not observed in their non-ideal MHD simulation.

\citet{2021A&A...652A..69M,2021A&A...656A..85M} performed radiation MHD collapse simulations also of a $100\:{M_\odot}$ core.
\cite{2021A&A...656A..85M} focused on the outflow.
They found mass outflow rates of $\sim10^{-5}-10^{-4}\:M_\odot\:{\rm yr^{-1}}$. 
The momentum rate that they found was $\sim10^{-4}\:M_\odot\:{\rm km\:s^{-1}\:yr^{-1}}$, which is much smaller than the $\sim10^{-3}-10^{-2}\:M_\odot\:{\rm km\:s^{-1}\:yr^{-1}}$ that we measure in our simulation. We also note that our model involves the momentum rate growing as the protostellar mass grows, while they found a roughly constant momentum rate with time. Also, contrary to our work, the opening angle in their simulations for the most part decreased with time.

\subsection{The role of the magnetic field}
\label{sub:magfieldrole}

\begin{figure*}[ht]
  \centering
	\includegraphics[width=0.9\textwidth]{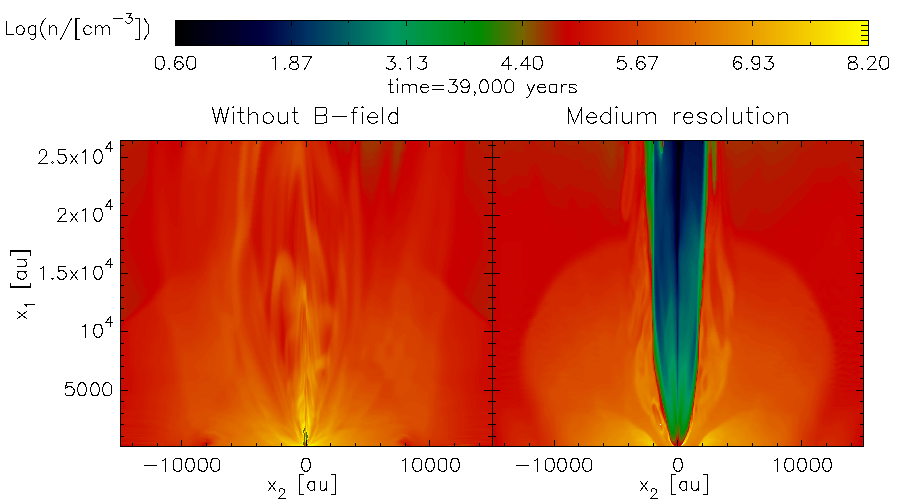}
    \includegraphics[width=0.9  \textwidth]{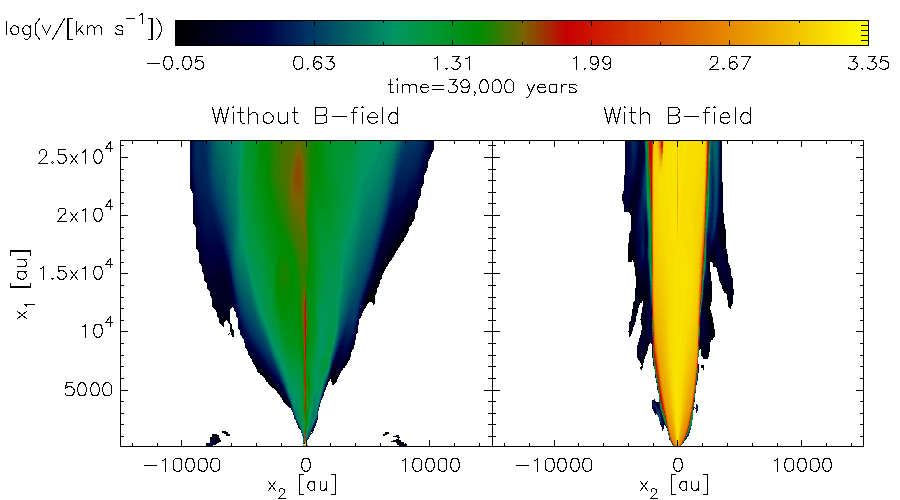}
	\caption{The effect of magnetic fields on the outflow structure is illustrated by a comparison of the number density in the $x_1 - x_2$ slice at $x_3 = 0$ and time 39,000 years, when the protostar is $8\:{M_\odot}$ for a case without magnetic field ($|B| = 0$) (left panels) and with a magnetic field (i.e., our fiduical model) (right panels). The upper panels show density structure; the lower panels show the velocity field of the outflowing gas. 
 }
	\label{bnob}
\end{figure*}

In ideal MHD, the gas is forced to follow the field lines. 
This therefore creates a natural separation between the outflowing gas and the collapsing envelope, because the field lines found in the outflow are anchored in the injection region.
To demonstrate this we performed a test simulation with the same set up, but without magnetic field.
In Fig.~\ref{bnob}, we show slices of the density structures and velocity fields of the outflowing gas for simulations with and without magnetic field after 39,000 years (i.e., when the protostar has reached $8\:{M_\odot}$).
A consequence of the lack of magnetic field is less collimated, slower outflow, which interacts with much more envelope material, causing a larger mass flow rate out of the simulation box as more envelope material is entrained in the outflow.
We also find that the outflow cavity is much less distinct, i.e., in its density contrast with the infall envelope, in the simulation without magnetic field.
Because of this, there is no high-velocity outflow, and the momentum flow rate at a height of 25,000 au is smaller than in the simulation with magnetic field.
Interestingly, the outflow pushes more material sideways when there is no magnetic field to confine it, forcing envelope material farther away from the protostar where the gravitational force is weaker, causing the envelope to collapse more slowly.
As a consequence, the envelope ``puffs up'' sideways in the no-magnetic field simulation, and at 39,000 years it extends beyond the side boundaries (see density panels in Fig.~\ref{bnob}).

\subsection{Effect of numerical resolution}
\label{sub:resolution}

\begin{figure*}[ht]
  \centering
	\includegraphics[width=0.83\textwidth]{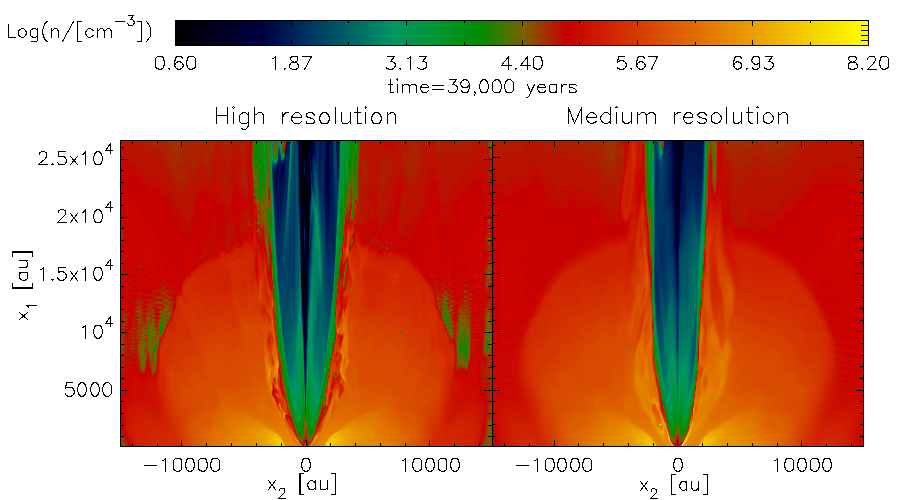}
 	\includegraphics[width=0.83\textwidth]{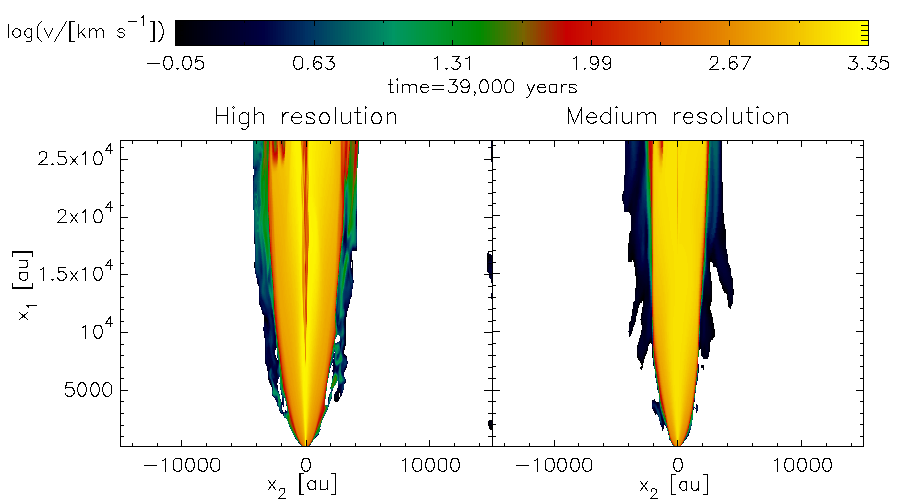}
	\caption{Effect of numerical resolution is illustrated by a comparison of the density structure in the $x_1 - x_2$ plane at $x_3 = 0$ at 39,000 years ($m_*=8\:{M_\odot}$) for the high resolution simulation (left panels) and fiducial medium resolution simulation (right panels). The upper panels show density structure; the lower panels show the velocity field of the outflowing gas.
 }
	\label{hrlrcomp}
\end{figure*}

To examine the dependence on numerical resolution, we ran the same simulation set up with twice as many cells in each direction (i.e., 336$\times$560$\times$560 cells; see \S\ref{sub:grid+bcs}), but keeping other parameters the same. 
In this higher resolution simulation, the smallest cells are now roughly 6 au on each side, compared to roughly 12 au in our primary ``medium'' resolution simulation.
This higher resolution simulation is much more computationally expensive, and it was not feasible to run it for the entire evolution (i.e., up to $\sim 24\:M_\odot$). Instead, we compare the results between the two resolutions at $t=39,000$ years, when the star has reached $8\:M_\odot$.
In Fig.~\ref{hrlrcomp}, we compare the logarithm of the number density, and the velocity field of the outflowing gas (where $v_1>0.9~{\rm km~s^{-1}}$), in a slice through the middle of the grid ($x_3 = 0$). 

The medium and high resolution simulations are qualitatively and quantitatively similar. For example, the opening angle of the outflow in the high resolution simulation measured at 12,000~au is $17.0^\circ$, compared to 
$20.0^\circ$ 
in the medium resolution simulation.
Note, while the low density part of the outflow cavity appears slightly larger in the slice of the high resolution simulation shown in Fig.~\ref{hrlrcomp}, the cavity defined by the outflowing gas is in fact slightly smaller. 
At 39,000 years, in the high resolution simulation we find that $1.5\:M_\odot$ has left the simulation box with the outflow through the outer $x_1$ boundary, while in the medium resolution simulation $1.2\:M_\odot$ has left the box.
These example diagnostics indicates a fairly good agreement between the higher and medium resolution simulations.

\section{Conclusions}
\label{summarysection}

We have presented a 3D-MHD simulation of a magnetically-powered disk wind outflow from a massive protostar located at the center of a core with initial mass of $60\:{M_\odot}$ and radius of 12,000~au. Such a core is the fiducial case of the Turbulent Core Accretion model of \citet{2003ApJ...585..850M}, which involves the core being pressure confined by an ambient clump medium with mass surface density of $\Sigma_{\rm cl}=1\:{\rm g\:cm}^{-2}$. In contrast to the ``snapshot'' approach of Paper I that investigated particular stages, we have followed the full, self-consistent evolution for 100,000 years as the protostar grows from $m_*=1\:{M_\odot}$ to about $26\:{M_\odot}$, following the protostellar evolutionary track of \citet{2014ApJ...788..166Z}, which sets both the accretion rate to the star and the mass and momentum injection rate to the disk wind outflow.

We find that the protostar drives a powerful, collimated outflow that breaks out of the core at relatively early times, i.e., within $\sim 1,000$~yr of the start of the simulation. At the scale of the initial core, the outflow has an opening angle (from outflow axis to cavity edge) of just over $10^\circ$ until $m_*=4\:M_\odot$ at 21,000~yr. Thereafter, as the protostar grows in mass and contracts towards the zero age main sequence, the outflow becomes more powerful causing the cavity to open up gradually, reaching opening angles of about $50^\circ$ by the end of the simulation. This disk wind outflow feedback thus dramatically affects the density structure and morphology of the protostar. While we have not performed radiative transfer (RT) calculations on these simulations (deferring this step for a future work), the RT models of \citet{2014ApJ...788..166Z} based on a semi-analytic core and outflow structure already illustrate the importance of such cavities for determining the infrared images and SEDs of the protostars.

The outflow also is the main factor determining the star formation efficiency (SFE) from the core. We find a lower limit to this SFE of $\bar{\epsilon}_{*f}=0.43$, but, considering the presence of a massive accretion disk and residual infall envelope, we estimate that the final value could reach as high as $\bar{\epsilon}_{*f}\simeq 0.7$. Such values are moderately higher than the efficiencies assumed of 0.5 in the fiducial TCA model of \citet{2003ApJ...585..850M}.

Inside the outflow cavity we find that the magnetic field is relatively weak, $\sim10^{-4}-10^{-5}~{\rm G}$, while it retains its initial core value $\sim10^{-3}~{\rm G}$ just outside the outflow cavity.
Near the base of the outflow, however, we find magnetic field strengths of $\sim0.1~{\rm G}$.
The magnetic field structure we have implemented acts to help separate the outflow from the collapsing core, limiting the amount of the envelope material being entrained in the outflow.

The mass flow and momentum rates of our simulation are $\sim2\times10^{-5}-2\times10^{-4}\:M_\odot\:{\rm yr^{-1}}$ and $\sim2\times10^{-3}-2\times10^{-2}\:M_\odot\:{\rm km~s^{-1}~yr^{-1}}$ respectively, with these values controlled by the boundary conditions we have implemented, but also comparable to rates measured from observed massive protostars. We have also compared the distribution of outflow mass with velocity, i.e., outflow mass spectra, of our simulations out to velocities of $\pm50\:{\rm km\:s}^{-1}$ with two example massive protostars G35.2 and G339 observed by ALMA. This comparison indicates that such observations have diagnostic power to constrain model parameters related to evolutionary stage, i.e., $m_*$, and viewing angle, i.e., $\theta_{\rm view}$. While precise agreement between model and observation is not found (and is not expected given potential systematic uncertainties in measure mass from CO line emission and from the limited range of TCA model parameters explored in our simulation), we do find quite striking agreement in the shape of the outflow mass spectra for some models. Further diagnostic tests involving full synthetic position-position-velocity cubes of synthetic CO line emission will be presented in a follow-up paper.

\begin{acknowledgements}

We would like to thank the referee for helpful comments and suggestions.
JES, JPR and JCT acknowledge support from Collaborative NSF grant AST-1910675. 
JES also acknowledges support from NASA through grant HST-AR-15053 from the
Space Telescope Science Institute, which is operated by AURA, Inc.,
under NASA contract NAS 5-26555.
JPR also acknowledges support from the Virginia Initiative on Cosmic Origins (VICO) and NASA grant 80NSSC20K0533.
JCT also acknowledges support from ERC Advanced Grant MSTAR.
We acknowledge the use of NASA High-End Computing (HEC) resources through the NASA Advanced Supercomputing (NAS) division at Ames Research Center to support this work. 
The analysis and the figures have been made using GDL \citep{2010ASPC..434..187C}, VisIt: https://visit-dav.github.io/visit-website/ , and Gnuplot: http://www.gnuplot.info/ .

\end{acknowledgements}

\bibliography{ms4}
\bibliographystyle{aasjournal}

\end{document}